\documentclass[sigconf]{acmart}
\usepackage{calligra}
\usepackage{algorithm}
\usepackage{algorithmicx}
\usepackage{algpseudocode} 
\usepackage{amsmath} 
\usepackage{booktabs}
\usepackage{subfigure}
\usepackage{multirow}
\usepackage{appendix}
\usepackage{paralist}
\usepackage{float}
\usepackage{svg}
\usepackage{algorithm}
\usepackage{algpseudocode}

\errorcontextlines\maxdimen

\makeatletter
\newcommand*{\algrule}[1][\algorithmicindent]{\makebox[#1][l]{\hspace*{.5em}\thealgruleextra\vrule height \thealgruleheight depth \thealgruledepth}}%
\newcommand*{\thealgruleextra}{}
\newcommand*{\thealgruleheight}{.75\baselineskip}
\newcommand*{\thealgruledepth}{.25\baselineskip}

\newcount\ALG@printindent@tempcnta
\def\ALG@printindent{%
	\ifnum \theALG@nested>0
	\ifx\ALG@text\ALG@x@notext
	\else
	\unskip
	\addvspace{-1pt}
	\ALG@printindent@tempcnta=1
	\loop
	\algrule[\csname ALG@ind@\the\ALG@printindent@tempcnta\endcsname]%
	\advance \ALG@printindent@tempcnta 1
	\ifnum \ALG@printindent@tempcnta<\numexpr\theALG@nested+1\relax
	\repeat
	\fi
	\fi
}%
\usepackage{etoolbox}
\patchcmd{\ALG@doentity}{\noindent\hskip\ALG@tlm}{\ALG@printindent}{}{\errmessage{failed to patch}}
\makeatother

\newbox\statebox
\newcommand{\myState}[1]{%
	\setbox\statebox=\vbox{#1}%
	\edef\thealgruleheight{\dimexpr \the\ht\statebox+1pt\relax}%
	\edef\thealgruledepth{\dimexpr \the\dp\statebox+1pt\relax}%
	\ifdim\thealgruleheight<.75\baselineskip
	\def\thealgruleheight{\dimexpr .75\baselineskip+1pt\relax}%
	\fi
	\ifdim\thealgruledepth<.25\baselineskip
	\def\thealgruledepth{\dimexpr .25\baselineskip+1pt\relax}%
	\fi
	\State #1%
	\def\thealgruleheight{\dimexpr .75\baselineskip+1pt\relax}%
	\def\thealgruledepth{\dimexpr .25\baselineskip+1pt\relax}%
}

\AtBeginDocument{%
  \providecommand\BibTeX{{%
    \normalfont B\kern-0.5em{\scshape i\kern-0.25em b}\kern-0.8em\TeX}}}

\copyrightyear{2023}
\acmYear{2023}
\setcopyright{acmlicensed}

\acmConference[MM '23] {Proceedings of the 31st ACM International Conference on Multimedia}{October 29--November 3, 2023}{Ottawa, ON, Canada.}
\acmBooktitle{Proceedings of the 31st ACM International Conference on Multimedia (MM '23), October 29--November 3, 2023, Ottawa, ON, Canada}
\acmPrice{15.00}
\acmISBN{979-8-4007-0108-5/23/10}
\acmDOI{10.1145/3581783.3613761}

\begin{document}

\title{MetaCast: A Self-Driven Metaverse Announcer Architecture Based on Quality of Experience Evaluation Model}


\author{Zhonghao Lin}
\affiliation{%
  \institution{The Chinese University of Hong Kong, Shenzhen}
  \city{Shenzhen}
  \state{Guangdong}
  \country{China}
}
 \email{zhonghaolin@link.cuhk.edu.cn}

\author{Haihan Duan}
\affiliation{%
  \institution{The Chinese University of Hong Kong, Shenzhen}
  \city{Shenzhen}
  \state{Guangdong}
  \country{China}
}
\email{haihanduan@link.cuhk.edu.cn}

\author{Jiaye Li}
\affiliation{%
  \institution{The Chinese University of Hong Kong, Shenzhen}
  \city{Shenzhen}
  \state{Guangdong}
  \country{China}
}
\email{jiayeli@link.cuhk.edu.cn}

\author{Xinyao Sun}
\affiliation{%
  \institution{Matrix Labs Inc} 
  \city{Vancouver}
  \state{BC}
  \country{Canada}
}
\affiliation{%
  \institution{University of Alberta} 
  \city{Edmonton}
  \state{AB}
  \country{Canada}
}

\email{asun@matrixlabs.org}

\author{Wei Cai}
 \authornote{Wei Cai is the corresponding author (caiwei@cuhk.edu.cn).}
\affiliation{%
  \institution{The Chinese University of Hong Kong, Shenzhen}
  \city{Shenzhen}
  \state{Guangdong}
  \country{China}
}
\email{caiwei@cuhk.edu.cn}

\renewcommand{\shortauthors}{Lin, Duan, Li, Sun and Cai}

\begin{abstract}
Metaverse provides users with a novel experience through immersive multimedia technologies. Along with the rapid user growth, numerous events bursting in the metaverse necessitate an announcer to help catch and monitor ongoing events. However, systems on the market primarily serve for esports competitions and rely on human directors, making it challenging to provide 24-hour delivery in the metaverse persistent world. To fill the blank, we proposed a three-stage architecture for metaverse announcers, which is designed to identify events, position cameras, and blend between shots. Based on the architecture, we introduced a Metaverse Announcer User Experience (MAUE) model to identify the factors affecting the users' Quality of Experience (QoE) from a human-centered perspective. 
In addition, we implemented \textit{MetaCast}, a practical self-driven metaverse announcer in a university campus metaverse prototype, to conduct user studies for MAUE model. 
The experimental results have effectively achieved satisfactory announcer settings that align with the preferences of most users, encompassing parameters such as video transition rate, repetition rate, importance threshold value, and image composition.
\end{abstract}

\ccsdesc[500]{Human-centered computing~Interactive systems and tools}
\ccsdesc[500]{Human-centered computing~Empirical studies in HCI}
\ccsdesc[500]{Human-centered computing~User studies}

\keywords{Metaverse, Announcer, Quality of Experience, Human-centered Computing, Virtual Cinematography}


\maketitle

\section{Introduction}
Metaverse is an emerging concept that is considered to be the next-generation Internet in which the users can completely immerse themselves in a 3D virtual space through interactive technologies. Currently, there are various directions of metaverse development, such as gaming-based (\textit{Roblox}, \textit{Minecraft}, \textit{Fortnite}, blockchain-based \cite{WeiCaiWEHFL2018} (\textit{Decentraland} and \textit{The Sandbox}), or extended reality-based \cite{metaverseQoE} such as \textit{Meta Horizon Worlds}. The common idea behind these projects is to provide a virtual world for users to interact with. As the metaverse attracts more users, the demand for an announcer system emerges \cite{aiannouncer}. Due to the sheer scale of the metaverse, users cannot cover it entirely with their own avatars. In this case, a metaverse announcer that broadcasts events in real-time can help users stay informed about ongoing events in the expansive virtual world.

However, most existing video game observers mainly serve in esports competitions, such as \textit{Counter-Strike: Global Offensive (CS: GO)}, \textit{League of Legends (LOL)}, and \textit{PlayerUnknown’s BattleGround (PUBG)}, etc. The esports live-streaming teams generally consist of a director and several observers. The observers monitor in-game data to keep track of ongoing events and capture highlight moments through a free camera. Subsequently, the directors select a video stream from all the observers’ captured streams and push it to the audiences \cite{JiayeLLWC2022}. However, as a parallel universe to the real world, the metaverse should be accessible anytime to anybody in the world, and so does its observer. This poses a challenge as human-driven observers are unable to afford 24-hour delivery in the practical metaverse. Currently, some researchers have developed virtual cinematographic systems to address specific needs, such as plot-based storytelling \cite{virtualstorytelling}, cinematic sequence \cite{kardan2008virtual}, and complex crowd tracking \cite{galvane2013steering}. Nonetheless, there is a lack of a general framework to fit the wide variety of situations occurring in the metaverse.

Therefore, in this work, we proposed a three-stage metaverse announcer architecture consisting of the Event Manager, Prose Storyboard Language (PSL) \cite{psl2013} Interpreter, and Camera Controller. The Event Manager captures newsworthy events from the metaverse, such as crucial user actions or crowd gatherings. After determining the targets to be announced, the PSL Interpreter will choose several shot specifications for the event and translate them to corresponding camera positions. At last, the Camera Controller takes charge of the path planning, including transition time and curve. 

Moreover, this paper identifies and analyzes various factors affecting the users' experience of a metaverse announcer, mainly containing two subjective factors: the accuracy of catching events and perceived video quality. For quantitative evaluation, we build a quality of experience (QoE) model, named Metaverse Announcer User Experience (MAUE) model, which comprises four objective factors: the transition time between shots, frequency of switching views, threshold value of important events, and image composition. 

To validate the proposed metaverse announcer architecture, we implemented an announcer system named \textit{MetaCast} in a university campus prototype. It can automatically capture events, film at an appropriate position, and drive the camera smoothly between shots. To determine suitable parameters for the MAUE model, we prepared video segments with different influencing factor settings and conducted user studies. Participants rated their satisfaction with the video segments through a 5-point Mean
Opinion Score (MOS). Based on our experimental data, we obtained a series of announcer system settings that align with most users' perceptions. 

Moreover, we also noticed that the results show significant variation among different users, e.g., someone prefers the bird's-eye view with a long duration, while someone prefers frequent switches in different events. To address this problem, we divided the experience into two stages. In \textit{S1}, \textit{MetaCast} adapts the settings from the previous experimental results. We invited 20 users to rate the performance, which shows that the selected parameter settings are palatable to most users. For \textit{S2}, we designed a real-time feedback system for users to express their preferences during their long-term daily use. The users' feedback demonstrates that \textit{MetaCast} could be subtly transformed to fit their preferences.

The major contributions of this paper are concluded as follows: 
1) we proposed a three-stage architecture of metaverse announcer, which fills the need of metaverse applications; 2) we identified factors affecting metaverse announcer users' experience and present a quantitative QoE evaluation model; 3) we implemented an automatic announcer system for a university campus metaverse prototype named \textit{MetaCast} based on the proposed three-stage architecture; 4) we conducted user studies to explore the factors affecting QoE and obtained a general setting for most users.


\section{Related Work}


\subsection{Intelligent Virtual Camera System}
Many researchers have attempted to build a self-driven intelligent camera system in virtual worlds \cite{bares1998realtime, camerabot,virtualcamcontrol}, and some of them integrated camera control into the gameplay module to provide a novel experience \cite{wang2021heraclitus,spacemaze}. For example, Halp et al. \cite{2001cameraengine} presented a camera engine for computer games. They emphasized controlling the camera by constraints and managed the tradeoff between constraint satisfaction and frame coherence. Hugh McCabe and James Kneafsey \cite{fpsCinematographysys} proposed a virtual cinematography system for first-person shooter (FPS) games. They applied a finite state machine (FSM) to choose different shots in particular situations. The seminal work \cite{christianson1996declarative} introduced the Declarative Camera Control Language (DCCL) as a general tool for describing idiom-based solutions for cinematography problems. Film idioms are solutions for obtaining good cinematography and editing in a range of predefined situations \cite{1996cinematographer,narrativesys}. After that, Ronfard et al. \cite{psl2013} introduced prose storyboard language (PSL) to define the syntax and semantics of a high-level shot description language. Based on PSL, Galvane et al. \cite{narrativesys} presented a novel importance-driven approach to generate cinematic replays for dialogue-based 3D video games rather than idiom-based ones. Due to the effectiveness of PSL, it will also be used in this paper to describe high-level shot specifications. On the other hand, Lino and Christie \cite{toricspace} dug into the area of camera control and proposed Toric Space, an efficient interval-based search technique for automated viewpoint computation of two or three targets. The technique has been reused in \cite{galvane2013steering} to create dedicated steering behaviors for automatically computing shots of crowd simulations. The studies mentioned above have significant contributions in their target areas and have partly inspired the camera control techniques in this paper, but there needs to be a framework to fit different metaverse. Therefore, this paper proposes a complete and general framework for metaverse announcers to fill the gap.

\subsection{Quality of Experience on Metaverse Projects}
With the popularity of the concept of the metaverse, one of many focuses of researchers has been to evaluate users' QoE in metaverse-related areas \cite{ZhonghaoLDWC2022,lessAnnoying,du2022rethinking}. For instance, as Porcu et al. \cite{metaverseQoE} discussed in their initial analysis of the QoE in the metaverse, human and system are the main factors potentially impacting the QoE of metaverse applications. When referring to virtual reality technologies, discomfort symptoms such as cybersickness is considered to evaluate the QoE. In \cite{wheelchairsimulator}, the author presented the results of an explicit (questionnaire-based) and implicit (physiology-based) evaluation of the immersive wheelchair simulator. Han et al. \cite{evaluatecinematographyinimmersive} conducted within-subject repeated-measures experiments in an immersive environment based on head-mounted displays (HMD), which explored the correlation between viewing experience and comprehensive factors. On the other hand, the system influential factors concern the network, infrastructure, hardware, and multimedia content. Schatz et al. \cite{OVsteaming} presented a first approach toward subjective QoE assessment for omnidirectional video streaming, including the stalling impact of HMD-based omnidirectional video and the difference between HMD-based and traditional television-based video with respect to stalling. Wang and Dey \cite{modelMGUE} analyzed the factors affecting the QoE of cloud server-based mobile gaming, including game genres, video encoding factors, and the wireless network condition. Moreover, they developed a prototype for real-time measurement of Mobile Gaming User Experience (MGUE) using controlled subjective testing. Nevertheless, the existing works have not conducted research on assessing the QoE of announcers in metaverse projects. Hence, this paper analyzes the factors potentially influencing users' perceived experience of metaverse announcers and presents the results of questionnaire-based measurement.

\section{Metaverse Announcer Architecture}
\begin{figure}[h!]
\vspace{-0.25cm}
\centerline{\includegraphics[width=\linewidth]{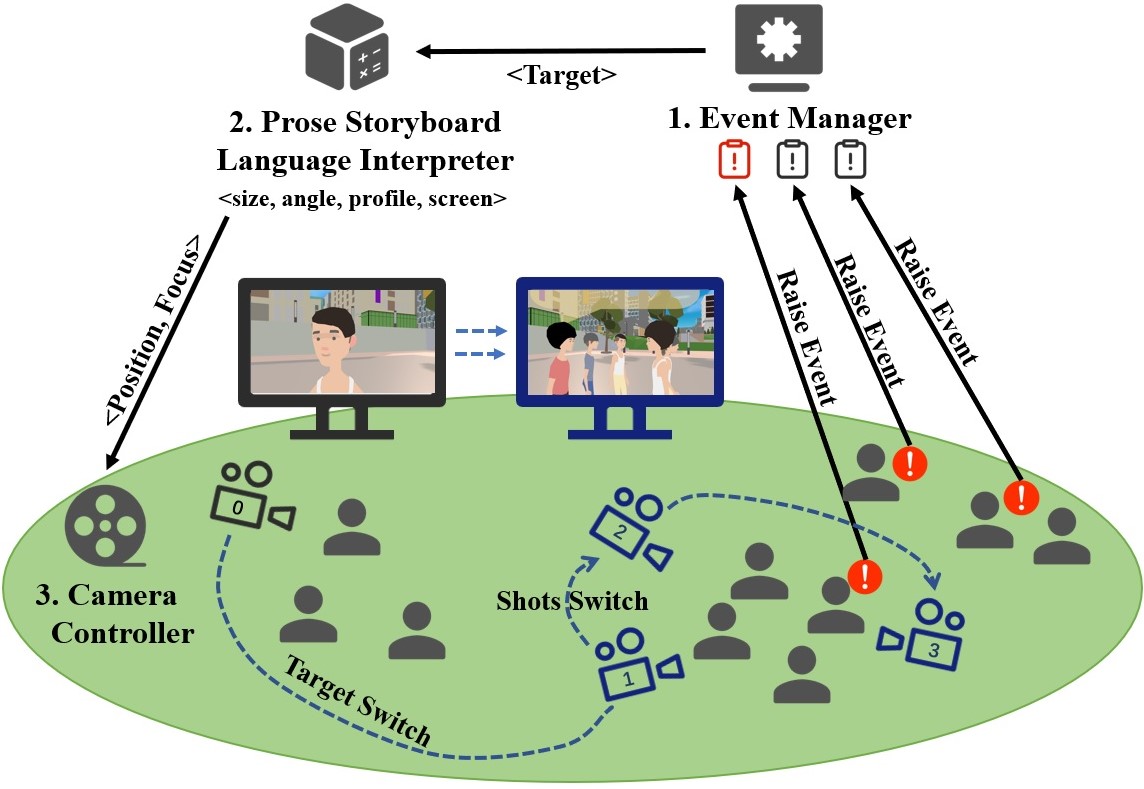}}
\caption{Metaverse Announcer System Architecture}
\vspace{-0.25cm}
\label{architecture}
\end{figure}

In this section, we present a three-stage architecture of metaverse announcers, including the composition and the overall workflow. As illustrated in Figure \ref{architecture}, the announcer is composited of three stages: Event Manager, PSL \cite{psl2013} Interpreter, and Camera Controller. The Event Manager is responsible for capturing ongoing events raised by users' behaviors. It will elect one event to announce and transmit details about the core avatars of the event to the PSL Interpreter. The PSL Interpreter selects several shot specifications and calculates the corresponding camera positions and rotations. Once the Camera Controller receives the target, it will move smoothly between each shot by the preset blend method with a suitable transition time.

\subsection{Event Manager}
The Event Manager is responsible for fetching important events both locally and globally in real time. Local events are those triggered by the significant actions of a few avatars. Since users' behavior can vary widely in different types of metaverse, prefabricated events triggered by specific users' actions such as McCabe et al.'s approach \cite{fpsCinematographysys} cannot fit our expectations as a general framework for metaverse announcers. Hence, we formulate the user's behavior as a function with different weights for different actions, 
\begin{equation*}
Importance = w_1 * a_1 + w_2*a_2 + ... + w_n*a_n (n>0)
\end{equation*}
where $a_n$ represents different actions and $w_n$ are corresponding weights for each action. 
As a general formulation, it allows designers to specify weights that are appropriate for their metaverse, such as assigning a high weight to User-Generated Contents (UGC) \cite{duan2022user} in gaming-based metaverses like \textit{Minecraft} and \textit{Roblox}. In blockchain-based metaverse like \textit{The Sandbox}, the transaction of Non-Fungible Tokens (NFT) \cite{chen2023information} should be regarded as an integral action while the token amount can partially determine the weight. 

Global events, on the other hand, are triggered by large gatherings of avatars in the virtual world. To monitor these events, we propose calculating the regional population density and applying an adaptive weight coefficient. By doing so, we can efficiently balance the ratio of local and global events announcements.


\subsection{Prose Storyboard Language Interpreter} \label{sec_psl_interpreter}
PSL is a high-level shot description language used to define the syntax and semantics \cite{psl2013}, which effectively inspires our architecture design. Table \ref{tab:psl} shows the subset of PSL we focus on. In this subset, <Subject> represents the target avatar of the shot. <Angle> denotes the relative height between the camera and the subject's face. While <Size> refers to the relative distance, from Extreme Closeup (ECU) to Extreme Long Shot (ELS). <Profile> expresses the relative position of the camera with respect to the subject. For example, <Profile = 3/4 right> means the camera is placed at 45 degrees to the right of the direction the subject is facing. <Screen> stands for the target avatar's position on the user's screen. 

\begin{table}[!h]
    \vspace{-0.2cm}
    \captionsetup{singlelinecheck=false}
    \caption{Simplified PSL Grammar (Shot Specification)}
    \vspace{-0.3cm}
    \begin{tabular}{l}
    \toprule 
    <Composition>::= <Angle, Size, Profile> on <Subject>[<Screen>]\\ 
    <Subject> ::= <Object> \\
    <Angle>::= Low | Eye | High \\
    <Size>::= ECU | BCU | CU | MCU | MS | LS | ELS \\
    <Profile>::= Front | 3/4 Right | Right | 3/4 Back Right | Back | \\
    \quad \quad \quad \quad \quad 3/4 Back Left | Left | 3/4 Left \\
    <Screen>::= Left | Center | Right \\
    \bottomrule 
    \end{tabular}
    \label{tab:psl}
    \vspace{-0.1cm}
\end{table}

Designers can specify the mapping table for different kinds of metaverse to customize shots. The high-level shot language can be transformed into camera positions and rotations. The input is the target avatar's position and the shot specification. It starts from the avatar's basic point, normally at its lower abdomen. Using <Angle>, we can determine the height of our target camera position. On this basis, we can get a directional vector defined by <Profile>. Move the camera at a distance determined by <Size> in this direction, and then we can get the camera position. The focus point begins from the midpoint of the target avatar's face, and the point should move a certain distance to the left or right according to <Screen> and <Size>. After that, control the camera to look at the focus point, then we can obtain the expected shot. 

When constructing a shot, there is a well-developed system in photography for us to follow. For example, the ``rule of thirds'' is a longstanding principle in photography and has proven to be useful \cite{proveruleofthirds}. It guides us to divide the frame up into 3*3 grids and place the focus point along one of the third lines or on one of the four intersections of the third line. Designers can adjust the mapping table to ensure most image compositions obey this rule. In addition, the ``look room principle'' \cite{cinematography-report} suggests giving extra room in the direction the subject is facing. Hence some specifications can be filtered out, such as <Profile=right, Screen=right>. Without giving enough looking room, the main avatars may look like they are talking to a wall, which is weird to the audience. After these filters, PSL Interpreter sends well-designed shots to the Camera Controller.

		

\subsection{Camera Controller}
The Camera Controller manages the camera's movement between shots, including the blend time and blend method. Similar to the shot filter, there are several basic rules in cinematography that guide camera control. For instance, during a 2-person dialogue, a filmmaker should establish the scene with a wider shot (e.g., long shot) containing both characters \cite{A-tool-for-computational-analysis-of-Narrative-Film,arijon1991grammar}, which is also known as the ``rule of introduction''. Moreover, the well-known ``180-degree rule'' states that the camera should remain on one side of the imaginary axis between two conversing actors to prevent viewer confusion and disorientation \cite{violate180degreerule}. By adhering to these rules, we can eliminate certain shot combinations. 

We assume three modes for metaverse announcers: bird's-eye view mode, first-person view, and third-person view \cite{Taylor2002VIDEOGP}. In normal cases, the camera moves as a dolly in preset traces to patrol around the metaverse and observe the digital world from a macro perspective. When global events happen, such as gathering crowds, the camera will move toward the nearest point on the trace and look at the target area to announce the event. When a local event occurs with only one target, the camera will switch to the first-person view of the target avatar, while local events with multiple targets, such as a conversation between several users, will trigger the third-person view mode with the shots decided by the PSL Interpreter.


\section{Quality of Experience Model}
In this section, we start by identifying and analyzing influential factors affecting the Metaverse Announcer User Experience (MAUE). After that, we developed a quantitative model with four perspectives to measure the perceived quality of users, named MAUE model.

\subsection{Influential Factors}
\begin{figure}[h!]
\centerline{\includegraphics[width=\linewidth]{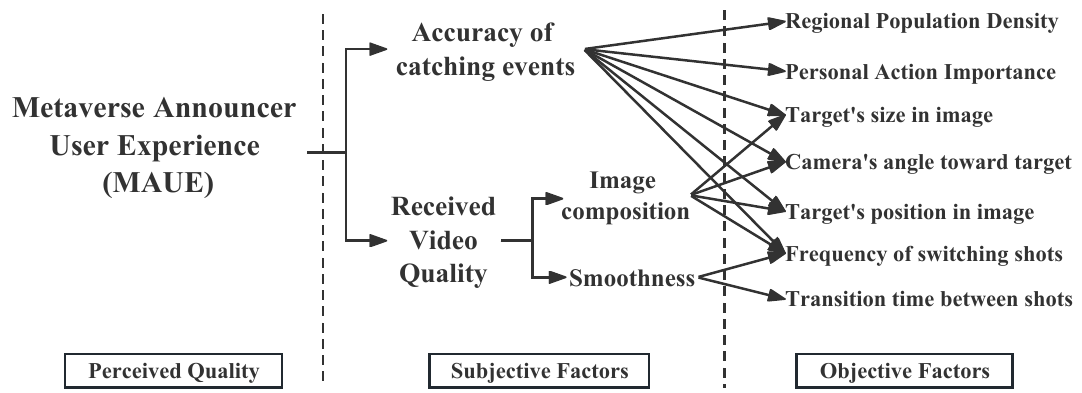}}
\caption{Objective and Subjective factors affecting MAUE}
\label{factors}
\end{figure}

Currently, existing work has discovered the QoE of network factors in mobile games, where poor network conditions will degrade the QoE \cite{modelMGUE}. Different from the existing work, our study aims to identify specific factors highly related to metaverse announcers rather than the influencing factors like network conditions (bandwidth, delay, etc.). We assume that metaverse announcers are ideal real-time systems with limited network delay. The most fundamental role of metaverse announcers is to precisely capture newsworthy events and broadcast them to the appropriate audience in time. Therefore, as depicted in Figure \ref{factors}, the MAUE model primarily depends on two subjective factors: the accuracy of catching events and the quality of received video content, determined by the image composition of each shot and the smoothness of switching views.

The subjective factors are determined by several objective factors. For instance, events are classified as global or local events. We use regional population density and personal action weighting functions to represent the importance, respectively. 
The image composition of shots can significantly influence users' perceived video content quality. We expect the camera to aesthetically capture the key elements of the event. Therefore, we quantify the image composition into three factors: the target's size, the camera's angle towards the target, and the target's position in the image. These factors can be represented by four parameters <Size>, <Angle>, <Profile>, and <Screen> based on PSL \cite{psl2013}. The smoothness of the video is determined by the frequency of switching views and the transition time between shots. Intuitively, an excessively low frequency leads to a boring experience, while a high frequency may cause dizziness and confusion. Similarly, the transition time between shots influences smoothness. Switching shots without any transition frame probably confuses the user, and users may lose focus when the transition time is too close to the shot duration. In Section \ref{sec_evaluation}, we will explore the parameter selection through quantitative user studies.

\subsection{Model Description}
The Mean Opinion Score (MOS), a subjective score ranging from 1 (unacceptable) to 5 (excellent), has been widely used to measure the QoE \cite{lessAnnoying, violate180degreerule}. In this paper, we adopt MOS to measure MAUE and develop a quantitative model. As shown in Figure \ref{factors}, MAUE depends on objective factors: the threshold importance value of raising events, image composition, transition time between shots, and frequency of switching views. Therefore, we formulate the MAUE model as follows:
\begin{equation*}
MAUE= f(Transition, Repetition, Importance, Composition)
\end{equation*}

\textbf{Transition.} In virtual environments, natural camera motion is crucial for a positive user experience. The transition between different camera shots is a significant factor influencing MAUE. The transition consists of several parts: transition method, curve, and time. The transition method refers to the trace of the camera's movement, such as moving linearly from point 1 to point 2, along with a spherical trajectory or a collision-free visible aware path proposed by \cite{visibilitytransitionplanning}. Violation of the ``180-degree rule'' will disorient viewers and weaken the user experience. Regarding the transition curve, various possibilities exist, such as linear, ease in hard out, hard in ease out, etc. In this paper, we primarily focus on exploring the influence of transition time on MAUE by conducting user studies to determine an appropriate transition time for metaverse announcers.

\textbf{Repetition.}
Repetition's impact on MAUE is evident in popular esports competitions, where directors alternate views to maintain suitable video repetition rates. For FPS games, first-person views of valuable players, battle replays, and the bird's-eye views of firefight areas \cite{JiayeLLWC2022} are common. In Multiplayer Online Battle Arena (MOBA) games such as \textit{League of Legends}, players control the actor from the isometric perspective \cite{lol}, and the in-game observing system can even provide a 3D view of highlight moments to the audience to decrease repetition, which is not accessible to the players themselves. This study quantifies repetition as view-switching frequency, seeking a balanced value for metaverse announcer users.

\textbf{Importance.} As a critical aspect of metaverse announcers, accurately and effectively capturing events, which impact MAUE, is tied to selecting an appropriate importance threshold value for raising events. A fixed value cannot handle diverse situations within the digital world. For instance, during a grand celebration event, the number of online users increases rapidly. An Event Manager with an importance threshold value designed for daily use will be triggered frequently since interactions between a large number of users generate a lot of events. Conversely, if a high threshold value is applied for daily use, no event deserves announcing to the users, which will lead to a dull experience. Therefore, the threshold value should be dynamically adjusted to handle diverse situations. 

\textbf{Composition.} Image composition influences the MAUE in two aspects: aesthetic composition improves the received video quality, and logical composition highlights the relationships between the characters, allowing users to grasp the key points at once. We adopt some basic photography rules discussed in Section \ref{sec_psl_interpreter} to avoid poor composition. After filtering out, approximately 150 kinds of image composition remain. However, it is hard to distinguish which one is better since they all obey the basic rules. Therefore, we adopt the 5-point MOS to collect participants' rank of the specifications. After that, we can build a database with the rank and score of each specification.

\section{System Implementation}

\subsection{Campus Metaverse Prototype}
CUHKSZ-Metaverse \cite{social-good} is a blockchain-driven university campus metaverse prototype developed by \textit{Unity} Engine, which has a low-poly style campus digital twin for our students to explore, interact, and conduct social activities. This metaverse establishes a connection with the real world, as the player's physical location influences the rate of acquiring digital tokens. The main gameplay is based on UGC creation \cite{duan2022crypto, duan2023metacube}, with players able to create voxel-style objects and mint them as NFT using the built-in editor. Overall, CUHKSZ-Metaverse exhibits most characteristics of a metaverse. 

\subsection{Details of \textit{MetaCast}}

We implemented a practical metaverse announcer demo called \textit{MetaCast} in CUHKSZ-Metaverse, following the architecture proposed in Figure \ref{architecture}. The Event Manager calculates the importance of each player's behavior every 10 seconds. Players' action in CUHKSZ-Metaverse mainly contains walking, running, jumping, talking, creating NFTs (pets and crafts) using the UGC editor, and NFT transactions. We evaluate the actions by the following factors: move distance, move speed, length of spoken words, number of voxels, and volume of NFT transaction. We attach corresponding weights to each factor respectively so that the importance of the player's action follows a normal distribution. In our blockchain-driven metaverse, the gameplay mainly focuses on UGC creation and NFT transactions \cite{duan2022crypto, duan2023metacube}, so their weights are higher than others. The importance threshold value for raising events dynamically adjusts based on the number of online users to ensure that the frequency of raising events is stable. After receiving the targets raised from the Event Manager, the PSL Interpreter will choose three shot specifications randomly, pass them to the photography rule filter, and calculates the camera's position and rotation. The Camera Controller drives the camera to the target position smoothly in a collision-free trace. Moreover, we preset the shot duration as 5 seconds and the blend curve as ease in and out. 

\begin{figure}[!h]
\vspace{-0.3cm}
    \centering
    \subfigure[First-Person View]
    {\includegraphics[width=0.32\linewidth]{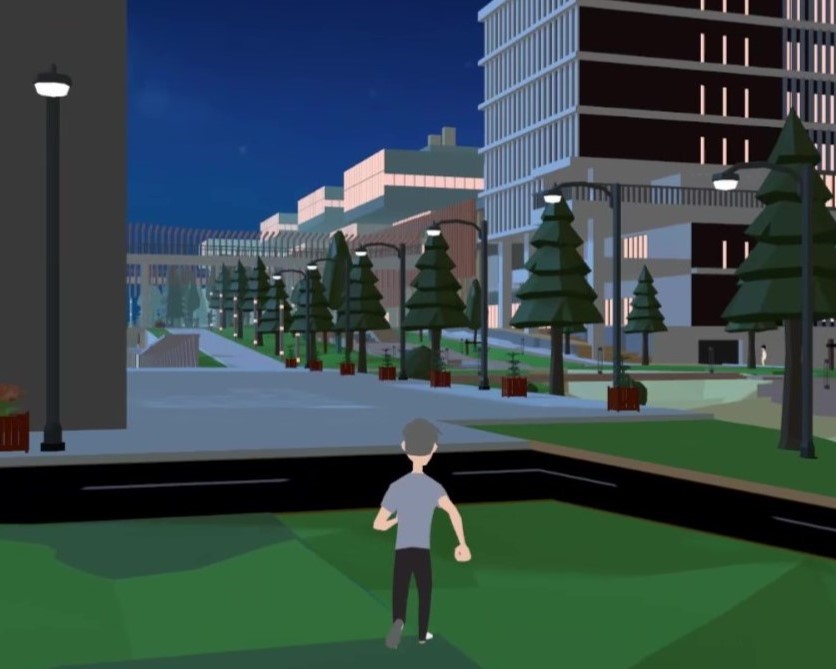}}
    \hfill
    \subfigure[Third-Person View]
    {\includegraphics[width=0.32\linewidth]{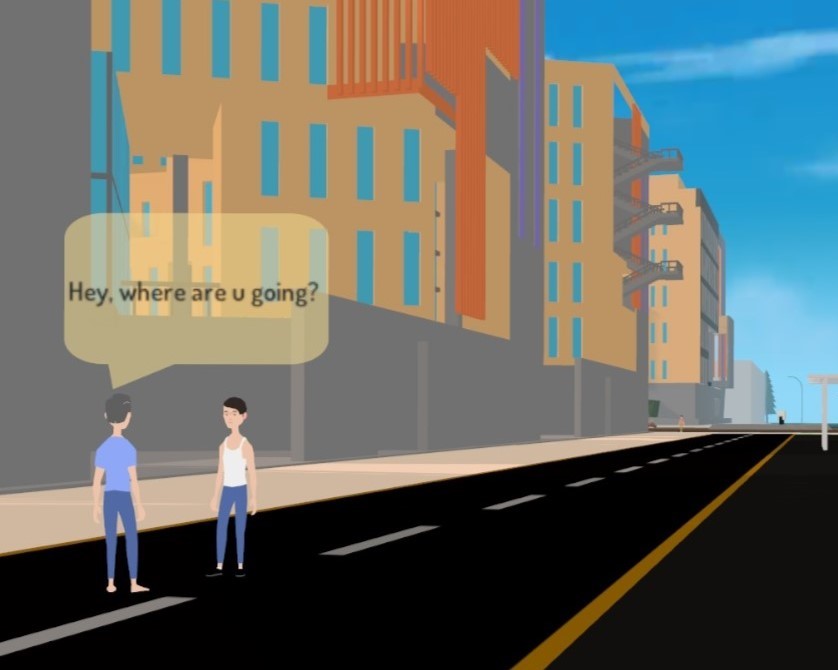}}
    \hfill
    \subfigure[Bird's-Eye View]{{\includegraphics[width=0.32\linewidth]{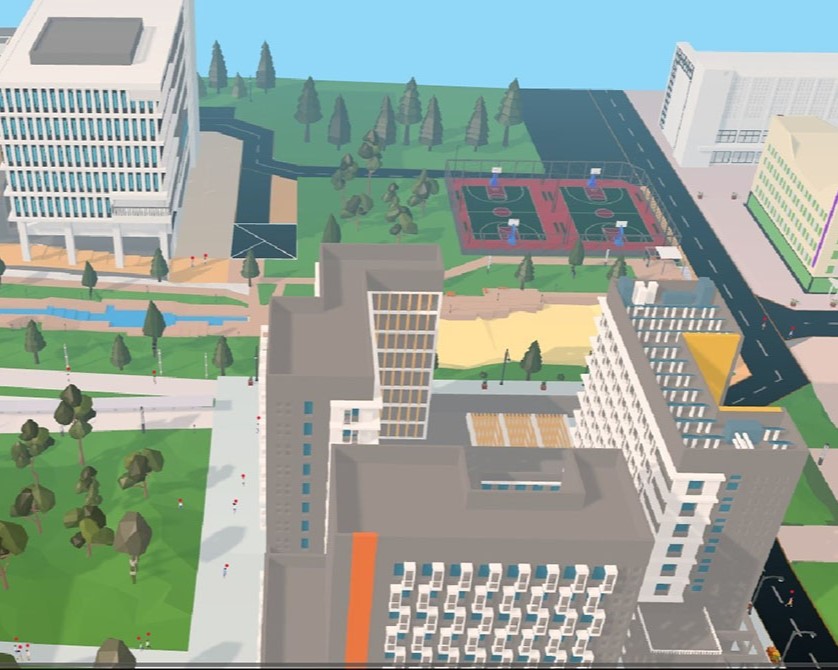}}\label{birdseye}}
\vspace{-0.5cm}
\caption{Three View Modes in MetaCast}
\vspace{-0.3cm}
\label{view modes}
\end{figure}

MetaCast offers three view modes: Bird's-Eye View, First-Person View, and Third-Person View modes, as illustrated in Figure \ref{view modes}. The Bird's-Eye View mode is designed for daily patrols and global events broadcast. As shown in Figure \ref{birdseye}, the camera observes the digital world from a bird's-eye view providing viewers with a macro picture, such as the distribution of users. Under normal circumstances, the camera moves as a dolly in a preset rounded trace, allowing viewers to observe the entire metaverse scene. During events, the camera switches to the first-person mode for a single target or the third-person mode for multiple targets, filming the event in sequence by randomly selecting three shot specifications.

It is important to note that the content presented through MetaCast is viewed on a 2D screen, and thus the experiment does not consider issues such as dizziness associated with head-mounted display (HMD) devices. This decision is based on the necessity of screen-based displays even in reality-based metaverse applications. For example, people in the real world may need to observe events happening within the metaverse through a large 2D screen. Moreover, metaverse players using HMDs can opt to observe activities in other areas through a virtual 2D display screen.

\section{Experiments} \label{sec_evaluation}

\subsection{Experimental Settings}
To maintain a stable experimental environment, we used behavior-tree-driven Non-Player Characters (NPCs) in the metaverse prototype. NPCs follow these steps: walk to a random destination, chase a random nearby NPC, have a conversation, wait 0 to 10 seconds, and choose a new destination randomly. By fixing the random seed value, we ensure that events in the metaverse remain consistent and reproducible. Hence, user studies can be conducted using controlled subjective testing. 

For the experimental participants, we recruited a total of 28 participants. Among them, twenty-five were young adults (two between 18-20 y.o. and twenty-three between 21-29 y.o.), followed by two participants between 30-39 y.o., and one between 40 and 49 y.o. The gender distribution is as follows: eight are female, and twenty are male, reflecting the current user distribution of CUHKSZ-Metaverse prototype, where most are college-aged students. 

\subsection{Evaluation on Transition} \label{sec_evaluation_of_transition}
 
\subsubsection{Experimental Setup.} 
\begin{figure}[h!]
\vspace{-0.5cm} 
    \subfigure[Video 1: a chat between two NPCs in the ``Start-Up Zone'']
    {\centerline{\includegraphics[width=\linewidth]{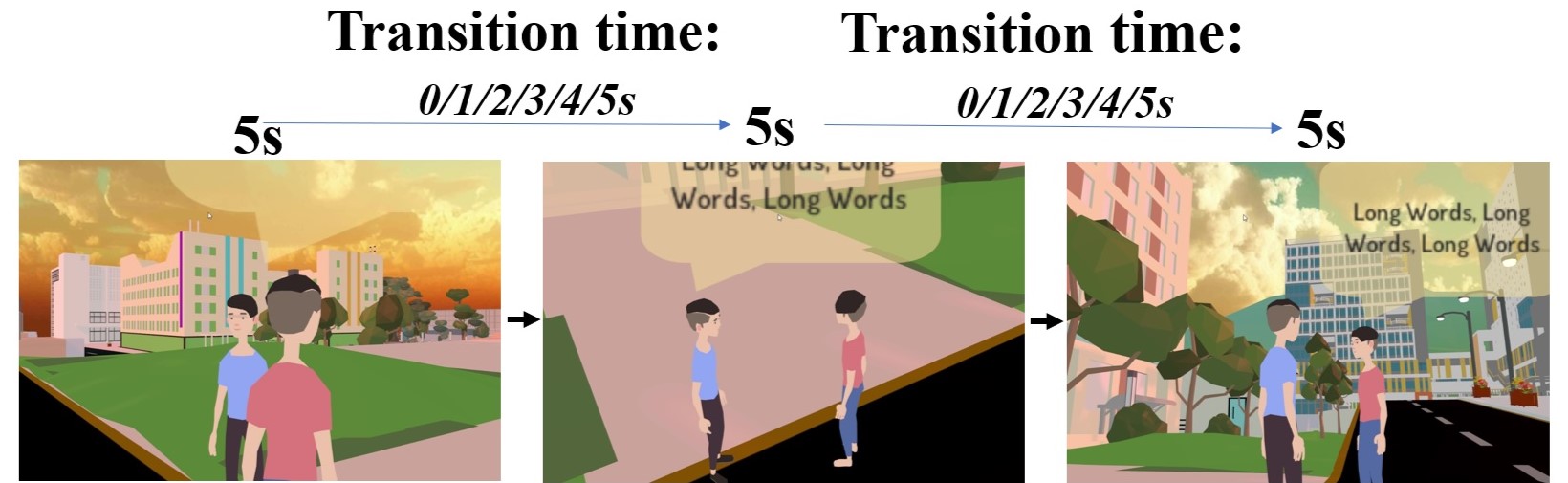}}}
    \subfigure[Video 2: a gathering of four NPCs in front of the ``Library'']
    {\centerline{\includegraphics[width=\linewidth]{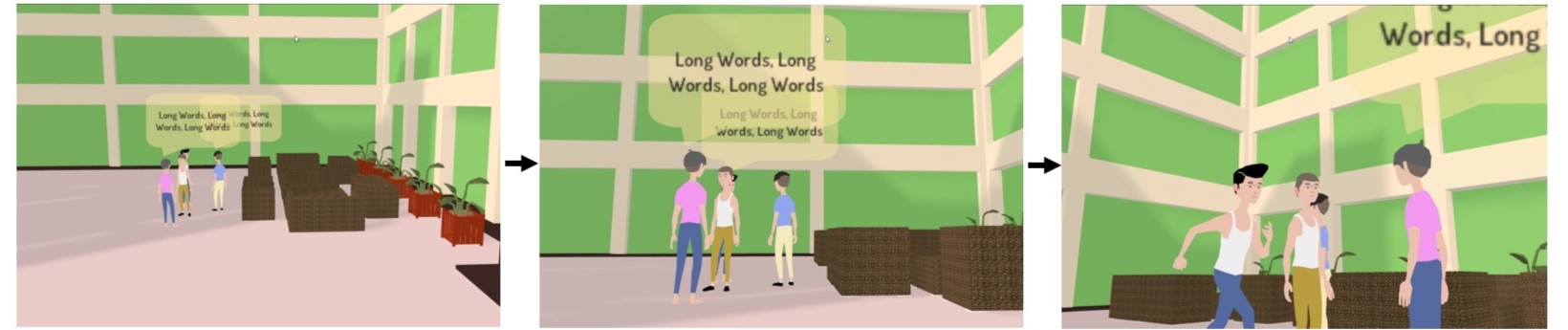}}}
    \subfigure[Video 3: a short chat of two NPCs by the ``Bus Station'']
    {\centerline{\includegraphics[width=\linewidth]{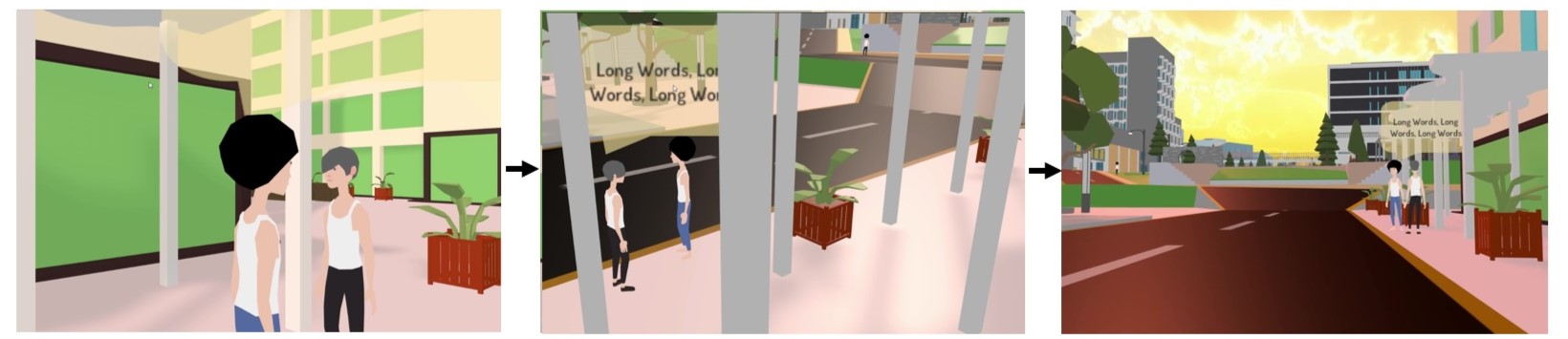}}}
\vspace{-0.5cm}    
\caption{Experiment Setting: Shots of Video 1, 2, 3}
\vspace{-0.25cm}
\label{transition}
\end{figure}

The video segment about a conversation between NPCs is used as the experimental material. As previous research \cite{A-tool-for-computational-analysis-of-Narrative-Film} found, the average duration time of a shot in animation genre films is 5 seconds. Since avatars in CUHKSZ-Metaverse are animation-style models with limited facial expressions, we classify the content of our metaverse announcer as animation and choose 5 seconds as the duration time of a shot. In a video plot, there are three 5-second shots with different PSL specifications. We changed the transition time between each shot to 0/1/2/3/4/5s to conduct the user study. We chose 5 seconds as the boundary condition since the transition time should not exceed the duration of the shot itself. As shown in Figure \ref{transition}, we chose three video plots filmed in different places of the CUHKSZ-Metaverse to avoid the influence of the image content.

There is a total of 18 videos (3 plots each with 6 groups of parameters) in the transition experiment. For the convenience of reading, we named them as Transition-plot-parameter. For instance, Transition-3-0 represents the 0-second transition version of plot 3. In simple terms, Transition-3-0 is a 15-second video segment, where each of the three frames in Figure \ref{transition}-Video3 stays for 5 seconds, and frame 1 will switch to frame 2 without any transition. For the user study, participants were first asked to watch Transition-1-n (0$\leq$ n$\leq$ 5) sequentially. They annotated the QoE level of their last watched video regarding the question ``How was your feeling about the transition speed between shots of the last watched video?'' using a 5-point MOS, which denotes the QoE from low to high: poor (1), bad (2), fair (3), good (4), and excellent (5). After that, they continued to watch and rank Transition-2-n (0$\leq$ n$\leq$ 5) with a two-minute rest between each plot to avoid the visual fatigue of the participants.

\subsubsection{Result and Analysis.} 
\begin{figure}[!h]
    \centering
    \subfigure[Average]
    {\includegraphics[width=0.48\linewidth]{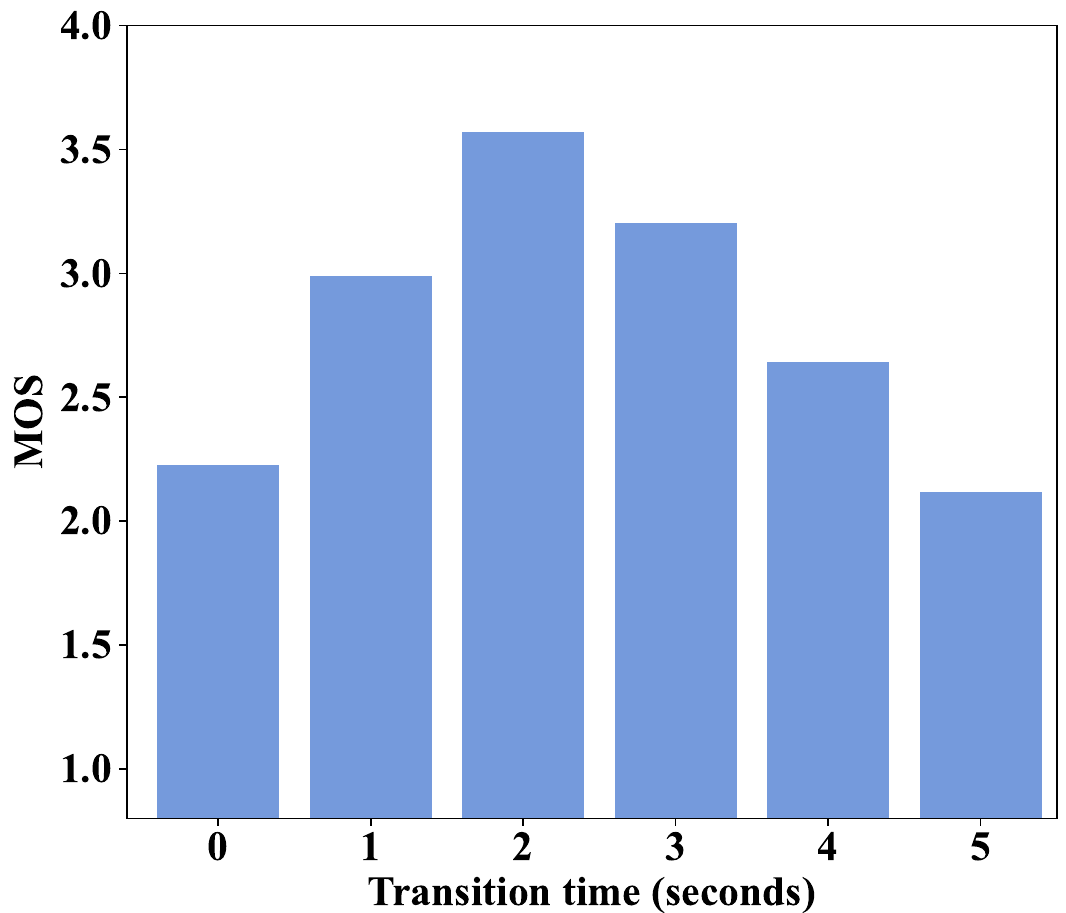}\label{trans-a}}
    \hfill
    \subfigure[Distribution]
    {\includegraphics[width=0.48\linewidth]{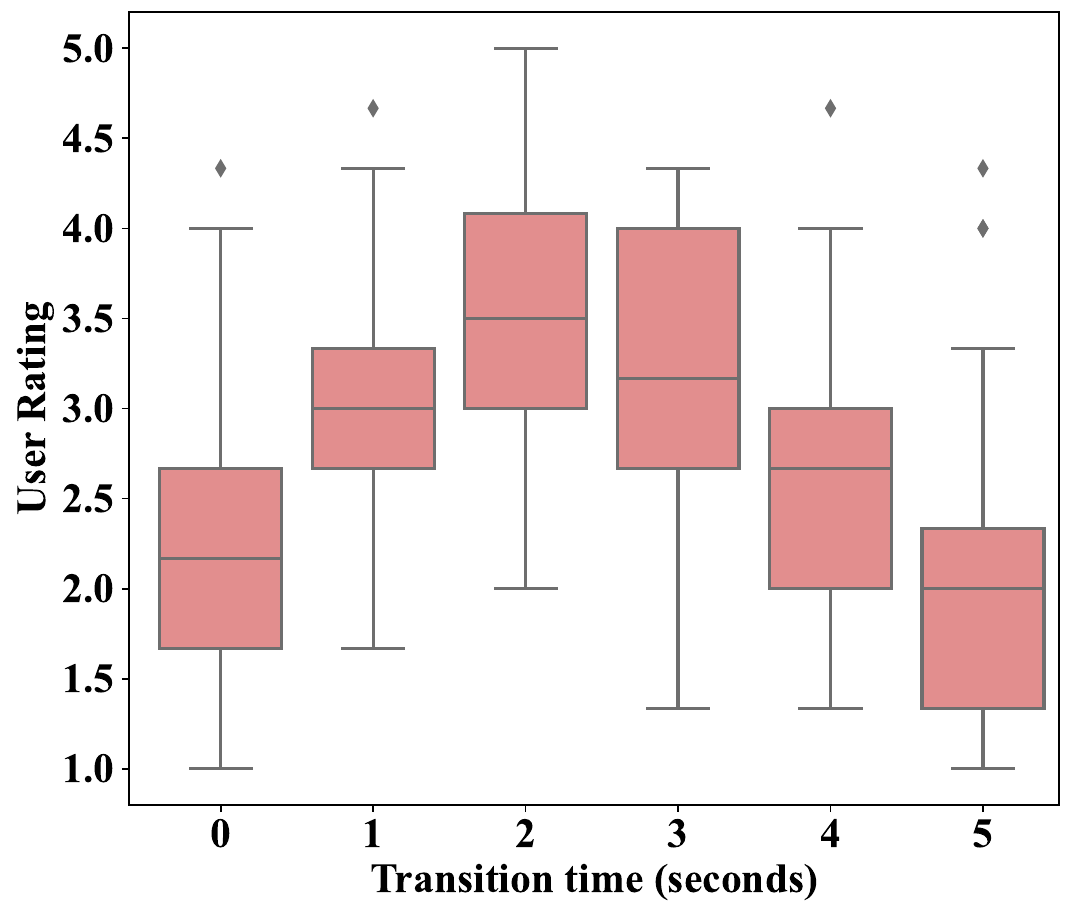}\label{trans-b}}
\vspace{-0.5cm}
\caption{Experiment Results on Transition}
\label{result: transition}
\end{figure}

We calculated the average MOS for each video and each participant's rating, as shown in Figure \ref{result: transition}.

Figure \ref{trans-a} displays the average score for different transition times, peaking at 2 seconds, while Figure \ref{trans-b} presents the distribution of user satisfaction, also showing the highest median score at 2 seconds. To ascertain that our results are efficacious, we performed Kruskal-Wallis Rank Test on the ordinal data, the result presented significant differences across the groups \textit{(H=93.152, p<0.01)}. Furthermore, we identified that Group Transition-2 had a significantly higher user rating than Transition-1 \textit{(Z=-3.733, p<0.01)} and Transition-3 \textit{(Z=-2.274, p<0.01)} by Wilcoxon Signed Rank Tests. For the 0-second cut, some participants mentioned that it cannot express the relation between shots, such as the relative position, where they even need to think about which angle the camera is shot from and how the camera moves from the last shot to the current shot. The 1-second version expressed the relative position of shots better, but the fast camera movement caused dizziness, leading to a decrease in users' quality of experience. Regarding the long transition time versions, they annoy the viewers due to the slow transition processes. The 5-second version scored lower than the 0-second version, indicating that most viewers preferred no blending over very slow blending between shots. In short, a 2:5 transition-to-shot-duration ratio offers the optimal QoE for metaverse announcer viewers.


\subsection{Evaluation on Repetition \& Importance}
In this subsection, we examine the impact of video repetition on users' QoE, quantifying video repetition as the frequency of view-switching in the metaverse announcer. A low view-switching frequency may result in monotonous content, causing visual boredom and aesthetic fatigue. To some extent, the importance threshold value determines the switching frequency: a low threshold value may bring more events to be announced, while a high value could result in no events being broadcast. Due to the strong relationship between repetition and importance, we combine them and conduct the experiment to explore the optimal view-switching frequency.

\subsubsection{Experimental Setup.}
Similar to the last experiment, we prepared three video segments with different plots by randomly setting the generation of NPCs, including number, spawn place, and suit. Each segment lasts for one minute, and we created six versions of each segment, each with a different frequency of view mode switching (1/2/3/4/5 times).
We chose 5 times as the boundary condition since we set the duration of events to be 10 seconds. Therefore, in that case, the camera is constantly broadcasting different events. For low-frequency cases, the camera remains in the bird's-eye view for most of the time. The content of the segments can be described as switching between view modes to observe different events as shown in Figure \ref{view modes}.
The video content is as follows:

\textbf{\textit{Video1}}: Bird's-Eye view (Start-up Zone) -> Two-avatar Conversation (University Gate) -> One-avatar Running (College A) -> Two-avatar Conversation (University Gate) -> Four-avatar Gathering (Library A) -> Four-avatar talking (Library B)

\textbf{\textit{Video2}}: Bird's-Eye view (Library A) -> Two-avatar Conversation (University Gate) -> Two-avatar Conversation (University Gate) -> One-avatar Walking (Start-up Zone) -> One-avatar Walking (College A) -> One-avatar Walking (Student Center)

\textbf{\textit{Video3}}: Bird's-Eye view (Teaching Buildings) -> Two-avatar Conversation (Library A) -> One-avatar Running (College A) -> Three-avatar Gathering (Teaching Building) -> One-avatar Walking (College A) -> One-avatar Running (Student Center) 

We applied the same form in Section \ref{sec_evaluation_of_transition} to name the video segment. The content of Repetition-3-2 is that the camera keeps the bird's-eye view mode for ten seconds and then switches to announce a two-avatar conversation at Library A. After that, it returns to the bird's-eye view for ten seconds and switches to the first-person view of the user running in front of College A. For the rest of the video, the camera patrols in bird's-eye view mode. The participants rated the QoE level of their last watched video through the MOS scale regarding the question ``How was your feeling about the frequency of switching views of the last watched video?''. A two-minute rest was set between each plot to avoid visual fatigue.

\subsubsection{Result and Analysis.} 
\begin{figure}[!h]
\vspace{-0.5cm}
    \centering
    \subfigure[Average]
    {\includegraphics[width=0.48\linewidth]{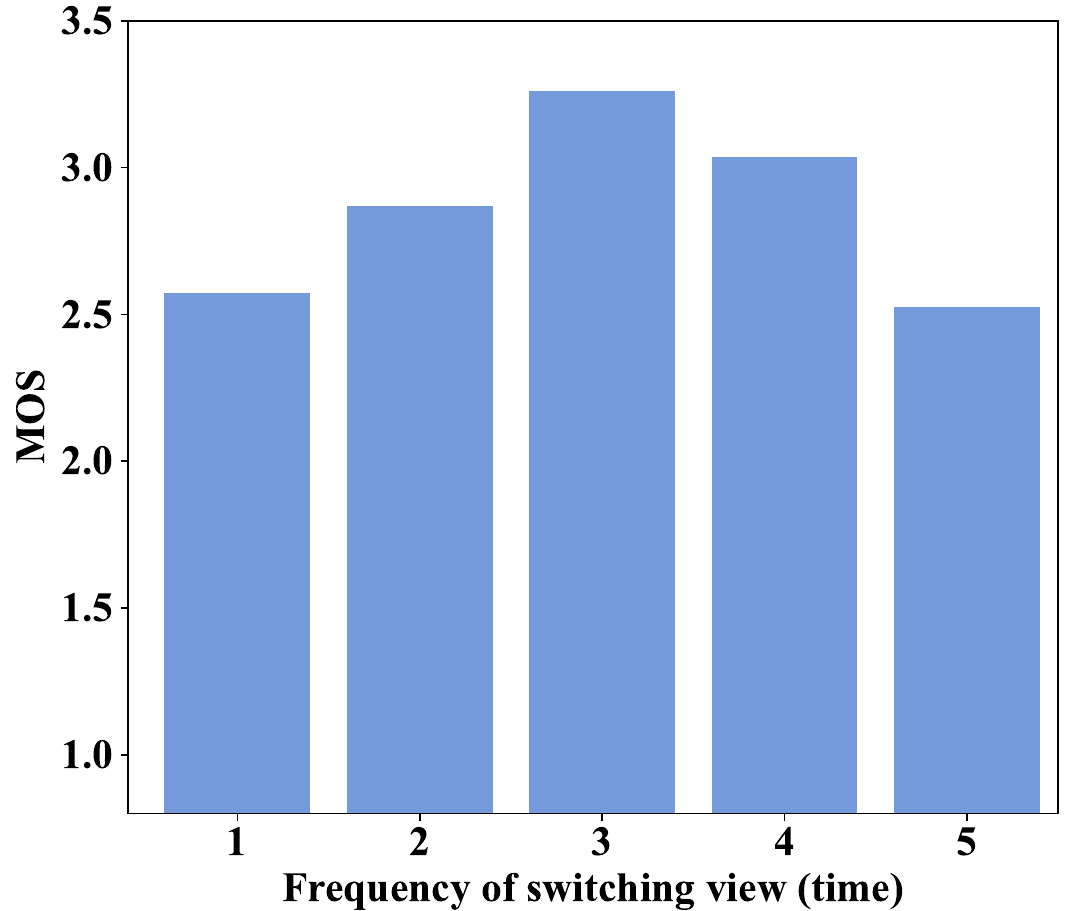}\label{repe-a}}
    \hfill
    \subfigure[Distribution]
    {\includegraphics[width=0.48\linewidth]{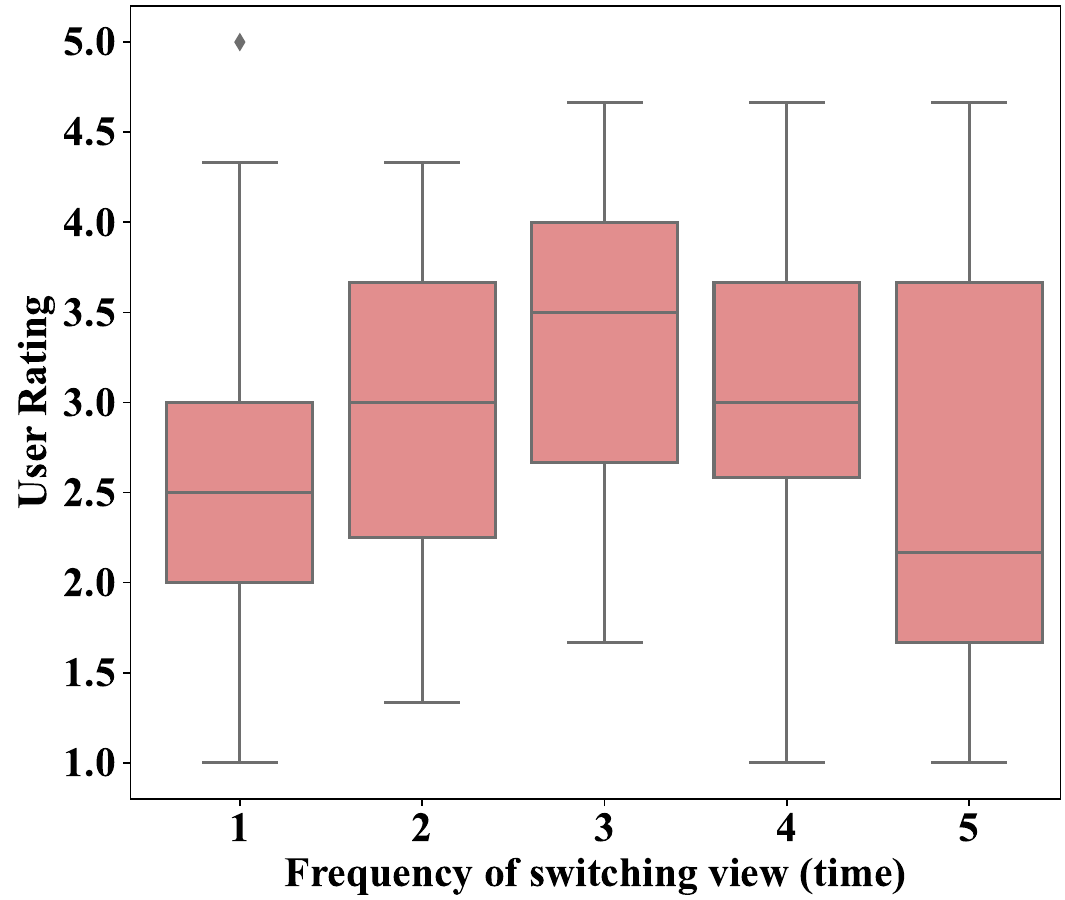}\label{repe-b}}
\vspace{-0.5cm}
\caption{Experiment Results on Repetition}
\vspace{-0.5cm}
\label{result: repetition}
\end{figure}

The results from the 28 participants are presented in Figure \ref{result: repetition}. 

\label{sec_variation}

Both the average and median scores indicate that, during a one-minute announcer, switching the view three times obtains the highest score in the experiment. To ascertain that our results are efficacious, we performed Kruskal-Wallis Rank Test on the ordinal data, and the result indicated significant differences across the groups \textit{(H=23.693, p<0.01)}. Additionally, we have identified that Group Repetition-3 has a significantly higher user rating than Repetition-2 \textit{(Z=-3.733, p<0.01)}, Repetition-4 has a significantly higher user rating than Repetition-5 \textit{(Z=-3.828, p<0.01)} by Wilcoxon Signed Rank Tests. However, we can observe the distribution of the MOS scores of the participants. For experiments with frequencies 3/4/5, the peak score is approximately 4.5, which implies that certain participants have a strong preference for these frequencies. In contrast, the minimum scores for frequencies 4 and 5 are as low as 1, indicating that some participants have a marked aversion to these frequencies. This reveals a considerable variation in participants' preferences for different frequencies. We conducted interviews with participants exhibiting diverse preferences. Some argued that during a brief period, there are few events that deserve the focus, leading them to favor the bird's-eye view of the world over focusing on unimportant events. While, a participant who assigned an average score of 4.67 to frequency 5 expressed that, compared to the monotonous bird's-eye view, he would rather explore what users are doing in various parts of the metaverse. Given the substantial preference gap among users, considering the median and lower bound, frequency 3 emerges as the optimal choice for most metaverse announcer users. In conclusion, the majority of users prefer metaverse announcers that allocate equal time to local and macro events, with a frequency that divides attention evenly between the two.

\subsubsection{Extension to Importance}
As mentioned above, the experiment video is organized as follows: a 60-second video, which catches events every ten seconds. Therefore, the result of frequency 3 can be interpreted in another way: most users prefer to use the metaverse announcers at the frequency that one out of every two fetch events can hit. Considering the difference in user behavior in each metaverse, we can generally assume the importance $I$ of users' behavior follows the normal distribution, $I \sim  N(\mu,\sigma^2)$.
In this case, the importance threshold value should be determined in real time according to the online user number. For instance, there are ten users currently in the metaverse. If we set the threshold value at 6.6\%, the probability that at least one of the ten users' behavior deserves announcing is 50.5\%. From the perspective of mathematical expectation, we can hit an important event every two fetches. The dynamic threshold percentage $i$ can be calculated by the number of online users $N$ and the frequency ratio of hitting event $f$ (our experiment result is $\frac{1}{2}$) by:
\begin{equation*}
     i = 1-\sqrt[N]{f},    0< i< 1, 0\leq f\leq 1
\end{equation*}
This algorithm is suitable for most kinds of metaverses, as long as the developer can implement the importance algorithm that conforms to the normal distribution.

\subsection{Evaluation on Composition}
\subsubsection{Experimental Setup.} Based on previous research \cite{A-tool-for-computational-analysis-of-Narrative-Film}, animated films tend to use MCU shots instead of CU due to the lack of detail in animated faces compared to human faces. Motivated by the study, we truncated ECU/BCU/CU on the <Size> parameter of PSL. And we had 3 <Angle>, 4 <Size>, 8 <Profile>, and 3 <Screen>, totally 288 combinations. As we mentioned in Section \ref{sec_psl_interpreter}, we used photographic rules such as look room \cite{cinematography-report} to filter out unsatisfactory shot specifications, resulting in 153 shots. Twelve participants were asked to rank the image composition of random specifications in this experiment. Each of them ranked 153 images organized by different specifications in different scenarios. Each time they wanted to switch to the next picture, they were asked to rate ``How was your feeling about the image composition of the last watched image?'' with a 5-point MOS scale.

\subsubsection{Result and Analysis.} 
\begin{figure}[htbp!]
\vspace{-0.5cm}
    \centering
    \subfigure[Eye-LS-Right-Left]
    {\includegraphics[width=0.48\linewidth]{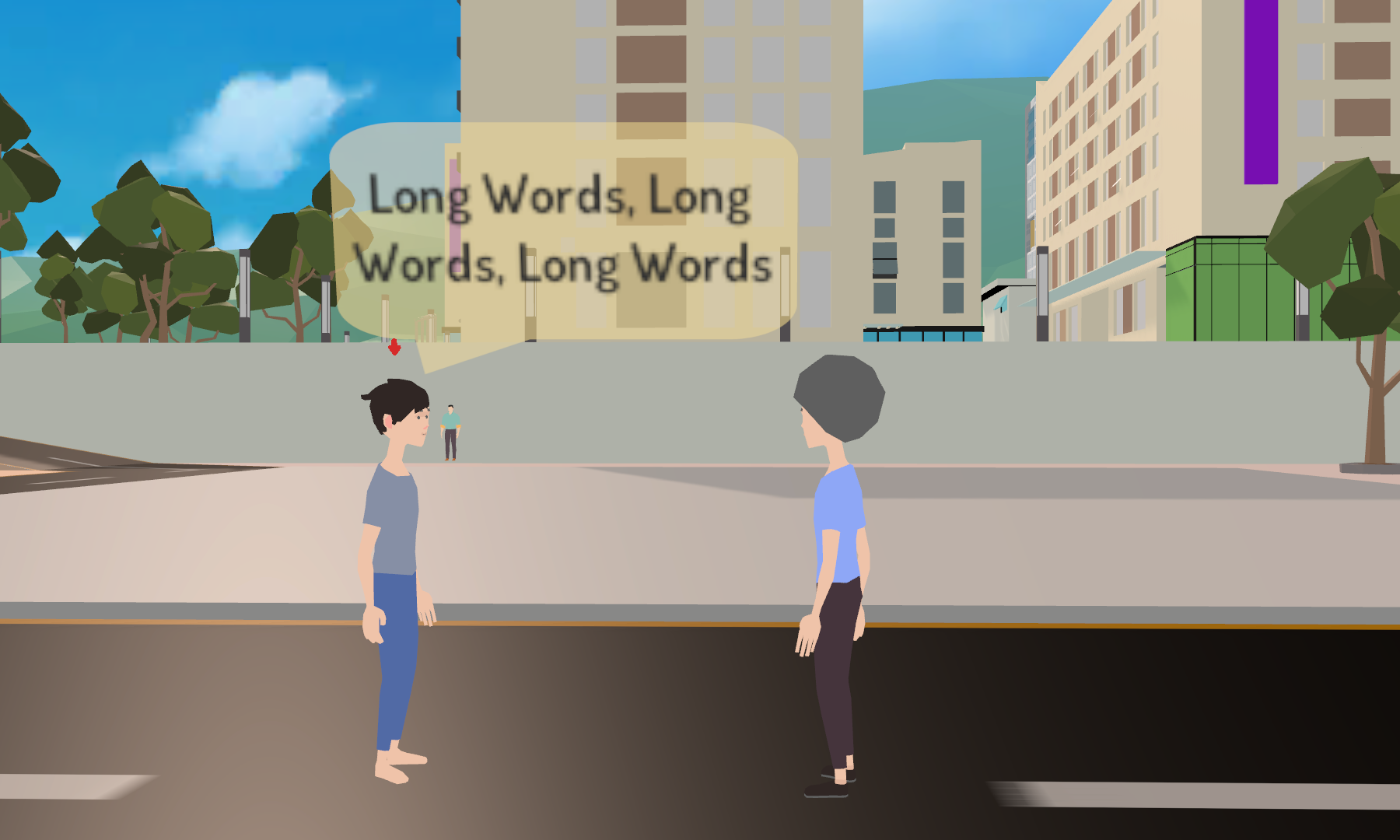}\label{7-a}}
    \hfill
    \subfigure[High-MS-Back-Left]
    {\includegraphics[width=0.48\linewidth]{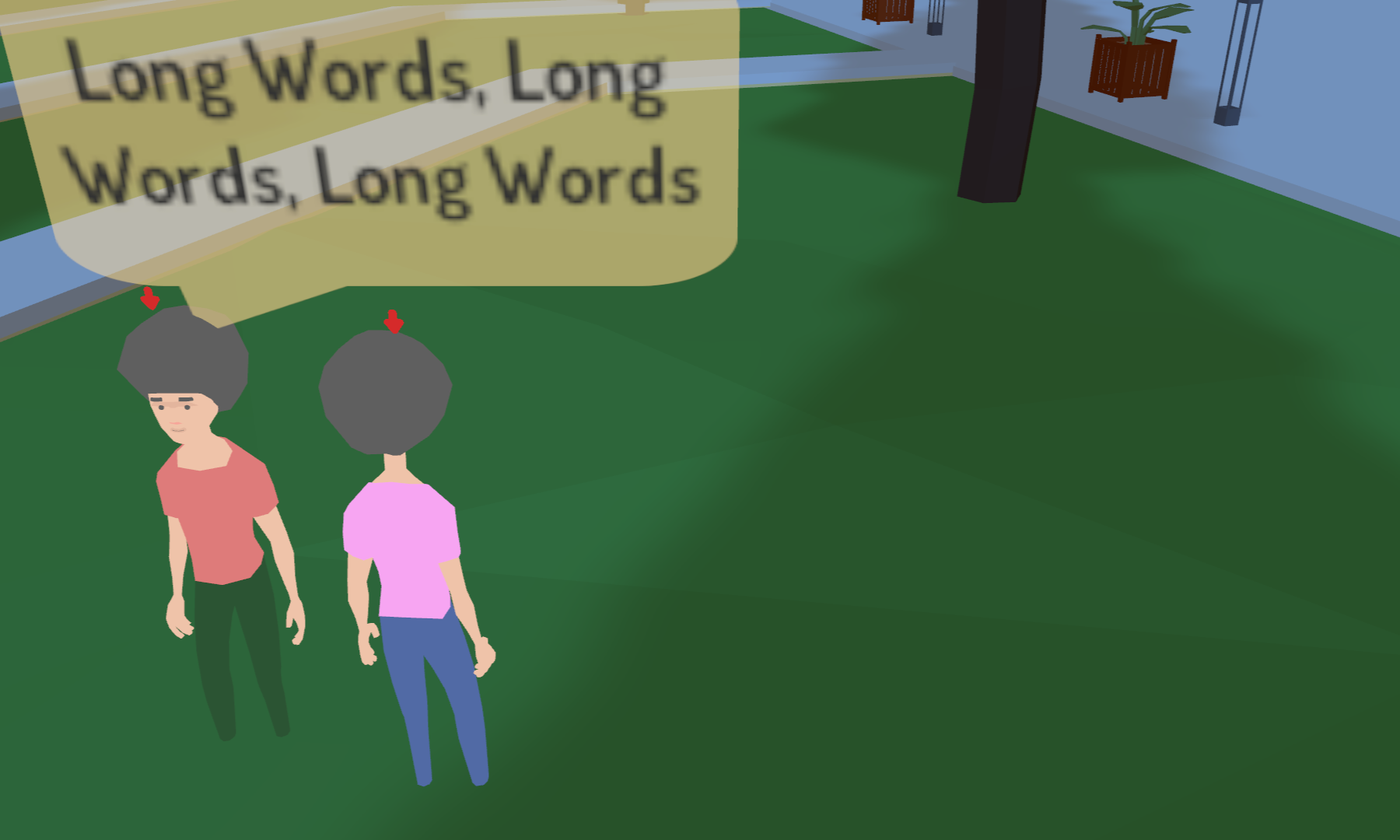}\label{7-b}}
\vspace{-0.5cm}
\caption{Image Composition Samples}
\vspace{-0.5cm}
\label{image composition}
\end{figure}

We collected 1836 ratings from 12 participants, as shown in Table \ref{tab:composition}. The Eye-LS-Right-Left specification received the highest rating, achieving a perfect score of 5. In simple terms, the camera takes a long shot at the subject's right side of the face while the subject is placed on the left side of the screen as illustrated in Figure \ref{image composition}(a).

We can find that it is a commonly used specification \cite{A-tool-for-computational-analysis-of-Narrative-Film, narrativesys} in movie filming for a two-person conversation: One person facing right is placed on the left side of the screen while the other facing left is positioned on the right side of the screen. The QoE ratings are mapped into two groups of high and low according to Schatz et al.'s user accessibility threshold \cite{poorOrGood}. Ratings higher than 3.5 \cite{lessAnnoying} are classified as the high group, while those below are assigned to the low group. After that, 42 specifications from rank 111 to rank 153 are classified as unsatisfactory image composition and filtered out. We identified several findings that may benefit future research: High-Level and MS (Medium Shot) is not a good combination as depicted in Figure \ref{7-b}. All the last three specifications have a combination of High-level and MS. Moreover, the highest MOS for any combination involving high-level and MS is 2.67, indicating that all of them are divided into unsatisfactory image composition.

\begin{table}[h!]
    \vspace{-0.2cm}
    \captionsetup{singlelinecheck=false}
    \caption{Top 5 Specifications of Image Composition}
    \centering
    \vspace{-0.3cm}
    \begin{tabular*}{\linewidth}{@{\extracolsep{\fill}}c|ccccc}
    \toprule 
    Rank & Angle & Size & Profile & Screen  &  MOS \\ 
    \midrule 
    1  & Eye&LS&Right&Left & 5 \\
    2  & Eye&MCU&34Right&Left & 4.83 \\
    3  & Low&LS&Left&Right & 4.67 \\
    4  & Eye&MS&Left&Right & 4.67 \\
    5  & Eye&LS&34Right&Left & 4.67 \\
    $\vdots$ & \multicolumn{5}{c}{$\vdots$}\\
    110 & High&ELS&34Leftback&Right & 3.5 \\
    $\vdots$ & \multicolumn{5}{c}{$\vdots$}\\
    151 & High&MS&Back&Right & 1.83 \\
    152 & High&MS&Front&Right & 1.83 \\
    153 & High&MS&Back&Left & 1.67 \\
    \bottomrule 
    \end{tabular*}
    \label{tab:composition}
\vspace{-0.5cm}
\end{table}

\subsection{Validation in \textit{MetaCast}}
\begin{figure}[htbp!]
\centerline{\includegraphics[width=\linewidth]{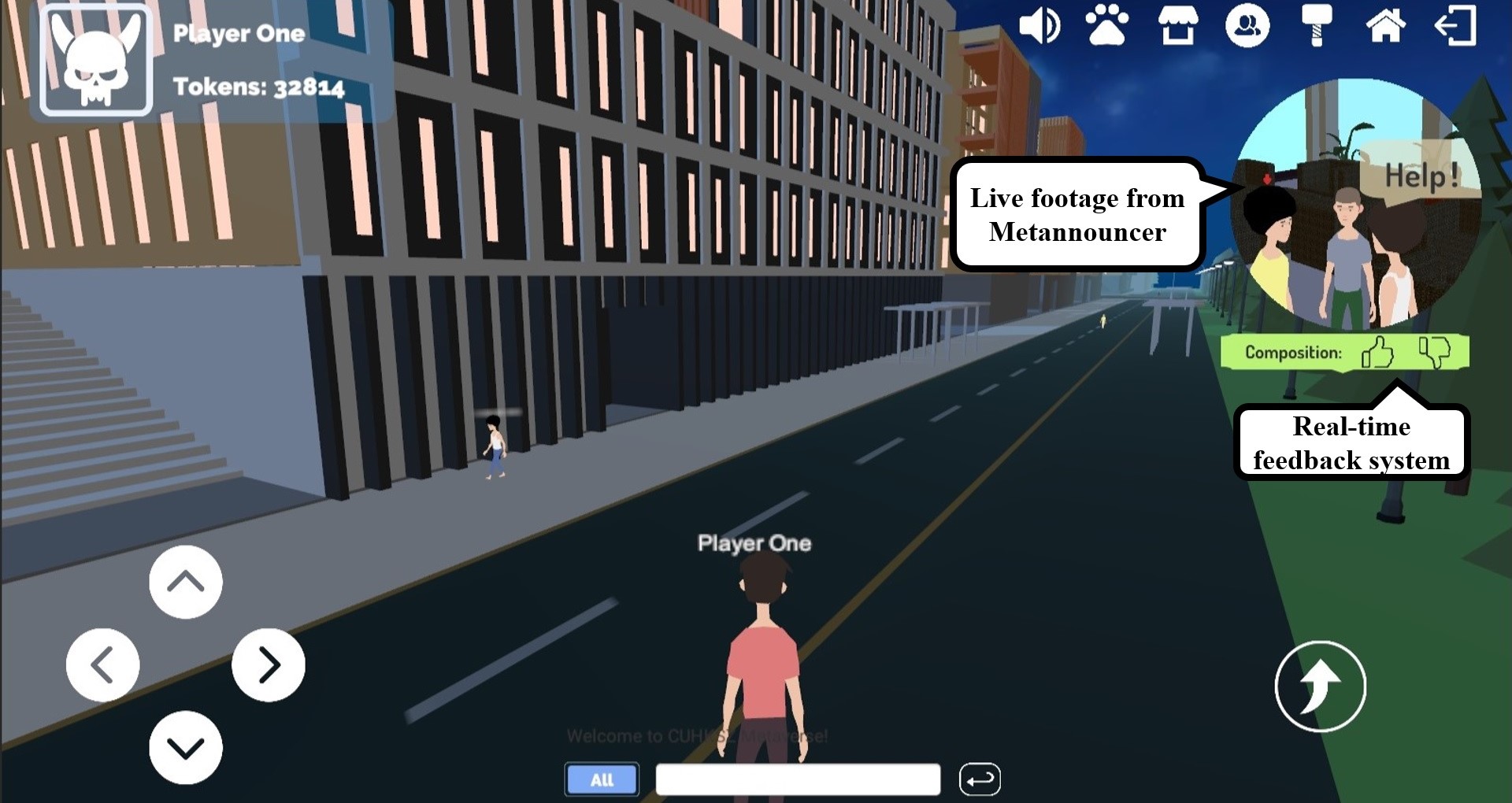}}
\vspace{-0.2cm}
\caption{Validation in CUHKSZ-Metaverse}
\vspace{-0.6cm}
\label{validation}
\end{figure}

In this section, we assess the performance of \textit{MetaCast} using a two-stage validation approach inspired by Alexandre et al. \cite{lessAnnoying}. The first stage, \textit{S1}, is to build a general model for most users using the results of the above experiments. As we mentioned in Section \ref{sec_variation}, we find that users' preferences can vary widely. Therefore, we have \textit{S2} to customize the announcer according to individual preferences through continuous feedback. This ensures a personalized metaverse announcer for each user. Moreover, \textit{MetaCast} is designed in two forms: (1) for on-metaverse users, when events happen, the video from the camera will be displayed in the top-right corner as shown in Figure \ref{validation}; (2) for offline viewers, \textit{MetaCast} provides a window for observing the digital world by screen-based displays.

During the \textit{S1} stage, the importance threshold value for raising events changes dynamically based on the number of online users to ensure that the frequency of raising events stays at 50\%. The PSL Interpreter will randomly choose three of the 110 specifications classified as the good group following Table \ref{tab:composition}. Then, the Camera Controller can drive the camera to the target position smoothly, and each shot stays for 5 seconds, and the transition time is set to 2 seconds. We invited 20 participants who have attended previous experiments to evaluate their overall feeling about the \textit{S1} stage of \textit{MetaCast} using MOS. We received an average score of 4.1 from the participants, which indicates a significant improvement compared to the previous experiments. It demonstrates the effectiveness of the proposed MAUE model.

In the player's daily use (\textit{S2}), the Ecological momentary assessment (EMA) \cite{EMA} system is triggered at each event. The users are asked to provide feedback with thumbs-up or thumbs-down emoji on the image composition as shown in Figure \ref{validation}. Similarly, the feedback for the transition time and the frequency of switching views is asked every ten minutes with a speed-up or slow-down emoji. The system remains unchanged if the user does not provide feedback. Otherwise, the value can be changed dynamically according to the user's feedback.

\section{Conclusion and Future work}
This paper presents a three-stage metaverse announcer architecture to automatically catch and broadcast events. We developed a quantitative MAUE model and conducted user studies to explore the relationship between factors and MAUE. Our findings reveal that a 2:5 transition-to-shot-duration ratio receives the highest user satisfaction. The majority of users prefer an equal balance between local and global events. The importance threshold for events should adjust dynamically based on the online user count. We also provided a table of ranked specifications for image composition. Consequently, our practical demo \textit{MetaCast} received positive user feedback, confirming the effectiveness of our proposed architecture.

The present study acknowledges certain limitations. On one hand, a larger, more diverse participant pool would strengthen the findings. On the other hand, the experimental conditions could be more rigorous; however, we opted for shorter experiments due to concerns about participant fatigue during longer video sessions. Moreover, a single-blind study in which participants are kept unaware of the experimental sequence is better. In future work, we intend to examine long-term feedback for metaverse announcer customization and explore AI-based approaches, including machine learning for event capture and classification, to enhance system performance. Furthermore, while the present study effectively enables the display of virtual 2D screens within HMD-based immersive environments, we aim to expand our scope, focusing on improving the integration of metaverse announcers and HMDs, addressing the challenge of dizziness induced by first-person perspective transitions during announcements, and ultimately, developing a more immersive metaverse announcer experience.

\begin{acks}
This work was supported by Shenzhen Science and Technology Program (Grant No. JCYJ20210324124205016), in part by the CUHK(SZ)-White Matrix Joint Metaverse Laboratory.
\end{acks}

\bibliographystyle{ACM-Reference-Format}
\bibliography{reference}

\end{document}